\documentclass[twocolumn,twocolappendix]{aastex63}

\usepackage{amsmath}
\usepackage{hyperref}

\newcommand\kms{km s$^{-1}$}
\newcommand\masyr{mas yr$^{-1}$}
\newcommand\teff{$T_{\rm eff}$}
\newcommand\logg{$\log g$}

\newcommand\msun{$M_\odot$}
\newcommand\lsun{$L_\odot$}
\newcommand\rsun{$R_\odot$}

\usepackage{listings}
\usepackage{CJKutf8}

\begin{document}
\title{2M17091769+3127589: a mass-transfer binary with an extreme mass ratio}
\begin{CJK*}{UTF8}{gbsn}
\author{Annaliese Miller}
\affil{Department of Physics and Astronomy, Western Washington University, 516 High St, Bellingham, WA 98225, USA}
\author[0000-0002-5365-1267]{Marina Kounkel}
\affil{Vanderbilt University, Department of Physics \& Astronomy, 6301 Stevenson Center Ln., Nashville, TN 37235, USA}
\email{marina.kounkel@vanderbilt.edu}
\affil{Department of Physics and Astronomy, Western Washington University, 516 High St, Bellingham, WA 98225, USA}
\author[0000-0001-9037-6180]{Meng Sun(孙萌)}
\affil{Department of Astronomy, University of Wisconsin-Madison, 475 North Charter Street, Madison, WI 53706, USA}
\author{Don Dixon} 
\affil{Vanderbilt University, Department of Physics \& Astronomy, 6301 Stevenson Center Ln., Nashville, TN 37235, USA}
\author{Chase Boggio}
\affil{Department of Physics and Astronomy, Western Washington University, 516 High St, Bellingham, WA 98225, USA}
\author[0000-0001-6914-7797]{K. R. Covey}
\affil{Department of Physics and Astronomy, Western Washington University, 516 High St, Bellingham, WA 98225, USA}
\author[0000-0002-3481-9052]{Keivan G.\ Stassun}
\affil{Vanderbilt University, Department of Physics \& Astronomy, 6301 Stevenson Center Ln., Nashville, TN 37235, USA}
\author[0000-0002-7130-2757]{Robert Mathieu}
\affil{Department of Astronomy, University of Wisconsin-Madison, 475 North Charter Street, Madison, WI 53706, USA}
\email{}

\begin{abstract}

We present the orbital solution of a peculiar double-lined spectroscopic and eclipsing binary system, 2M17091769+3127589. This solution was obtained by a simultaneous fit of both APOGEE radial velocities and TESS and ASAS-SN light curves to determine masses and radii. This system consists of an $M=0.256^{+0.010}_{-0.006}$ \msun, $R=3.961^{+0.049}_{-0.032}$ $R_{\odot}$ red giant and a hotter $M=1.518 ^{+0.057}_{-0.031}$ \msun, $R=2.608^{+0.034}_{-0.321}$ $R_{\odot}$ subgiant. Modelling with the MESA evolutionary codes indicates that the system likely formed 5.26 Gyrs ago, with a $M=1.2$ \msun\ primary that is now the system's red giant and a $M=1.11$ \msun\ secondary that is now a more massive subgiant. Due to Roche-lobe overflow as the primary ascends the red giant branch, the more evolved ``primary'' (i.e., originally the more massive star of the pair) is now only one-sixth as massive as the ``secondary''. Such a difference between the initial and the current mass ratio is one of the most extreme detected so far. Evolutionary modelling suggests the system is still engaged in mass transfer, at a rate of $\dot{M} \sim 10^{-9}$ \msun / yr, and it provides an example of a less evolved precursor to some of the systems that consist of white dwarfs and blue stragglers.

\end{abstract}

\keywords{}

\section{Introduction} 
\label{sec:intro}

Modern stellar evolution theory dictates that the time it takes for a star to evolve off of the main sequence is strongly dependent on its initial mass. However, in direct conflict with this, observations have discovered close binary stars where the more evolved stellar component is also the less massive of the pair. These systems are commonly referred to as Algol binaries \citep{Peter2001} after the classical example, and are most commonly thought to be created via Case B Roche-lobe overflow \citep{Tout1988}. The evolutionary destination of Algol binaries is heavily affected by the rate and efficiency of their mass-transfer history. Studies of Algol binaries support that their mass transfer (MT) can be significantly non-conservative, though the mechanism for how mass escapes the system is poorly constrained by observations \citep{Deschamps2015}. Furthermore, attempts to model Algol binary evolution with non-conservative MT struggle to reproduce the mass ratios of the observed population \citep{Rensbergen2011}.

As a byproduct, Roche-lobe overflow evolution of Algol binaries acts as a formation channel to several classes of post mass-transfer stars, which include blue straggler stars, carbon-enhanced metal-poor stars, barium stars, et cetera \citep{Jorissen2019}. Detailed modeling of the main-sequence components of Algol binaries is useful to constrain epochs and rates of non-conservative MT \citep{Sun2021}. Algol systems also prove interesting as potential progenitors of close white dwarf binaries, which are highly anticipated gravitational wave sources that could be detected by the future LISA space observatory \citep{Lu2020}. 

Here we present an orbital solution for 2M17091769+3127589 (here after 2M1709), a newly identified double-lined eclipsing binary consisting of a red giant star and a subgiant star. This system was previously identified as a Cepheid variable with a pulsation period of 2.933 days \citep{akerlof2000}, but \citet{price-whelan2018} and \citet{pawlak2019} subsequently identified the system as a 5.866 day binary, based on their analyses of APOGEE-derived radial velocities (RVs) and ASAS-SN light curves, respectively. 
We present in \ref{sec:data} the Gaia, APOGEE and ASAS-SN data recorded for 2M1709, along with a recently obtained TESS light curve. We report a joint radial-velocity and light-curve fit (Sect. \ref{sec:jointfit}) using the PHOEBE eclipsing binary modeling code and a best-fit MESA model (Sect. \ref{sec:mesa}) to investigate the system's mass-transfer history and eventual fate. With such careful definition, 2M1709 can serve as an excellent benchmark system for better understanding Algol binary evolution.

\section{Data}\label{sec:data}

\subsection{Gaia}

We use astrometry from the Gaia EDR3 data release \citep{gaia-collaboration2020} to determine the distance and space motion of 2M1709. Gaia reports a parallax of $1.014\pm0.015$ mas, $\mu_{\alpha} = 7.143 \pm 0.014$ \masyr, $\mu_{\delta} = -18.749 \pm 0.016$ \masyr. EDR3 also reports 2M1709's Renormalized Unit Weight Error (RUWE) as 1.013, well below the threshold of 1.4 established by \citet{lindegren2018} to identify systems with excess astrometric errors. The correlation derived by \citet{stassun2021} to predict photocenter semi-major axes for eclipsing binaries with RUWE $<$ 1.4 implies a photometric semi-major axis of 0.136 mas, or 13\% of the astrometric parallax reported in EDR3 for the system. This 13\% signal provides a conservative upper limit to the excess distance error that could be induced by the system's orbital motion, but it is likely a significant overestimate: converting the semi-major axis of our best fit orbit to an angular scale returns a value of 0.07 mas, well below the $>$0.1 mas threshold for which the \citet{stassun2021} relation applies. This physical semi-major axis also likely overpredicts the photocenter motion by a factor of several, as the optically brighter component is significantly more massive, such that the photocenter is likely quite stable. As a result, we conclude that the 1\% parallax uncertainty and low RUWE reported by EDR3 suggest 2M1709's distance is indeed within a few percent of 986 pc.

Combining the Gaia EDR3 astrometric solution with the center-of-mass RV of the system (Section \ref{sec:jointfit}), we estimate its Galactic velocity components to be U = $-82.6\pm0.3$, V = $-21.4\pm0.1$, and W = $-46.4\pm0.5$ \kms. These kinematics place the system between the thin disk and thick disk populations in a Toomre diagram \citep[e.g.,][]{bensby2005, reddy2006, nissen2009}.

\begin{figure}
 \centering
 \epsscale{1.3}
 \plotone{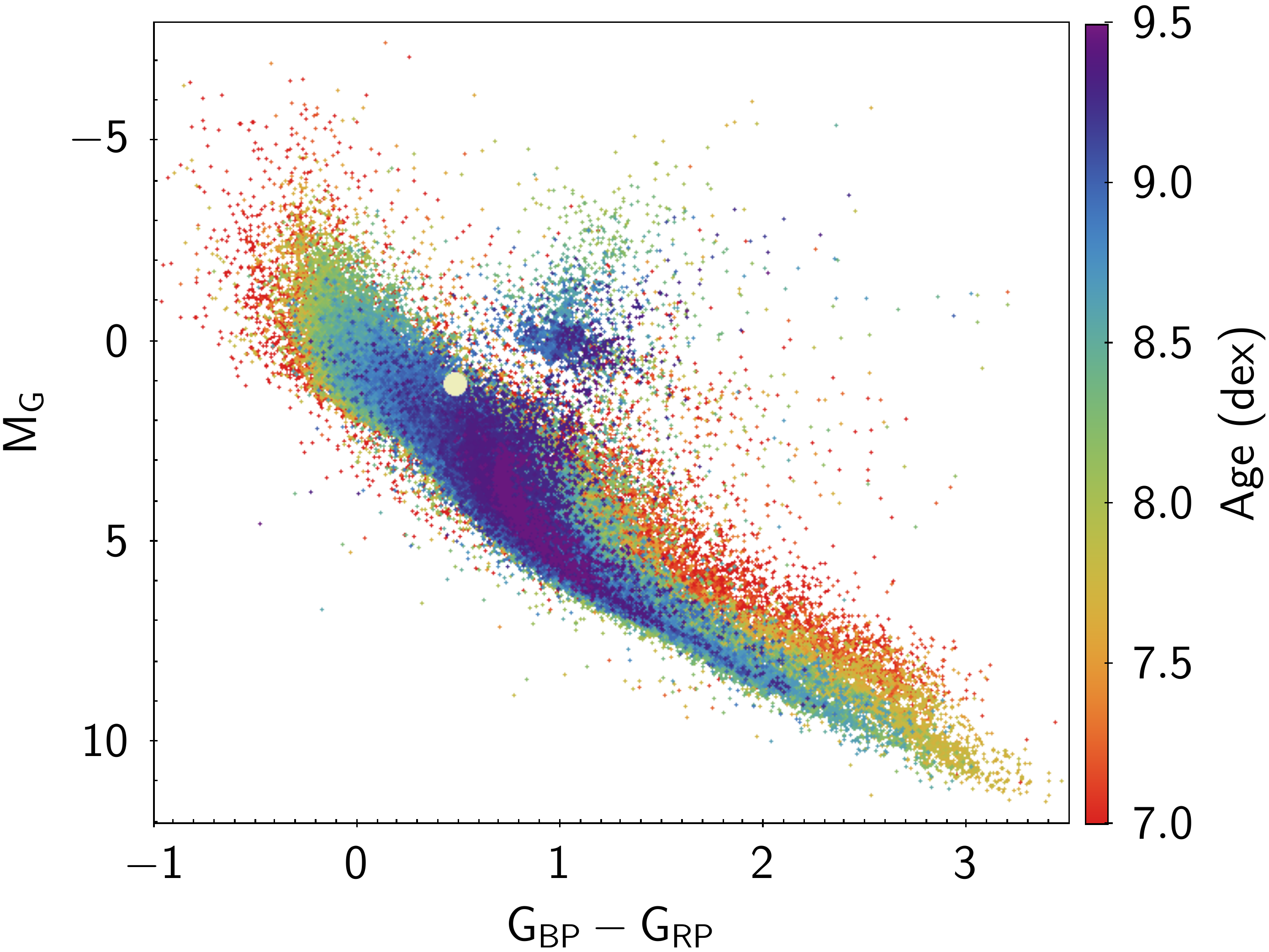}
 \caption{HR diagram showing the position of 2M1709 (large yellow circle), plotted against the background of nearby open clusters from \citet{cantat-gaudin2018a} color coded by their ages from \citet{kounkel2020}.}
 \label{fig:cmd}
\end{figure}

The system's optical photometry is most consistent with a source that is somewhat evolved off the main sequence (i.e, a hot subgiant; see Figure \ref{fig:cmd}). Gaia DR2 reports a photometric \teff\ estimate of $6424\pm90$ K, luminosity of $19.2\pm0.7$ \lsun, and a radius solution (assuming a single star) of $3.54\pm0.09$ \rsun \citep{gaia-collaboration2018}. Similarly, the source has been reported to have \teff\ of 7315 K in the optical LAMOST spectra \citep{luo2019}.

\subsection{APOGEE}

\begin{figure*}[!t]
 \centering
		\gridline{\fig{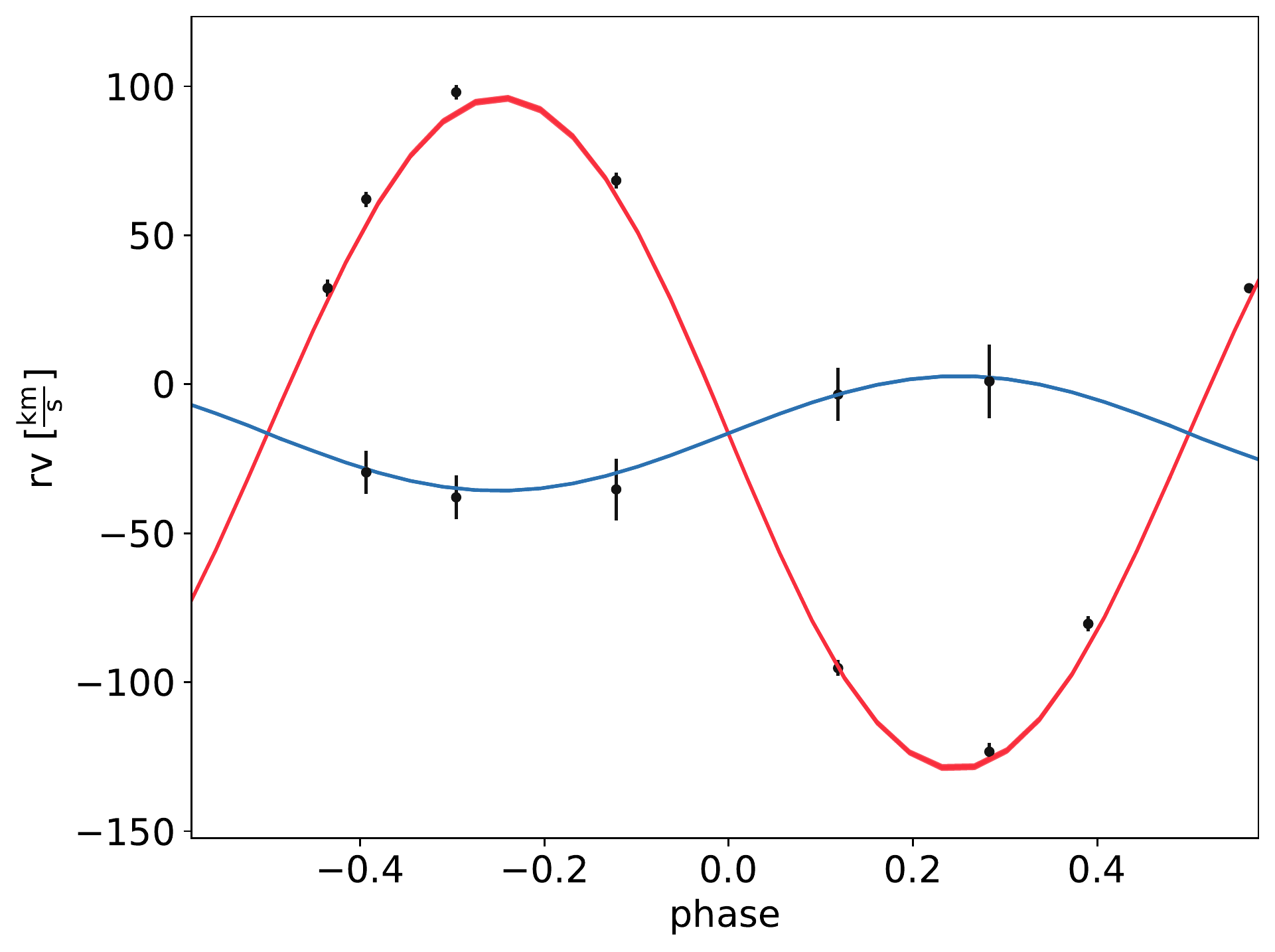}{0.5\textwidth}{}
 \fig{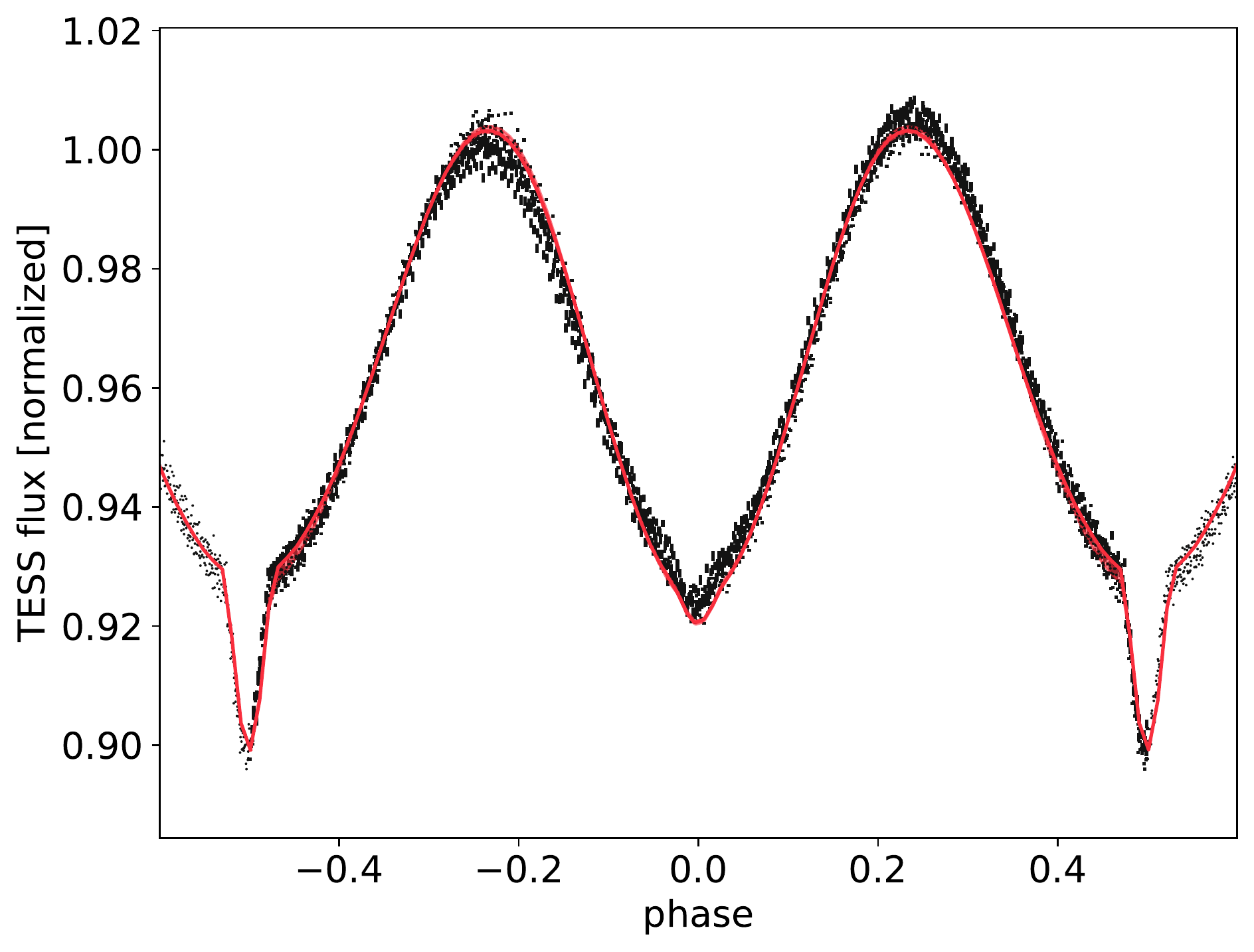}{0.5\textwidth}{}
 } \vspace{-0.8cm}
 \gridline{\fig{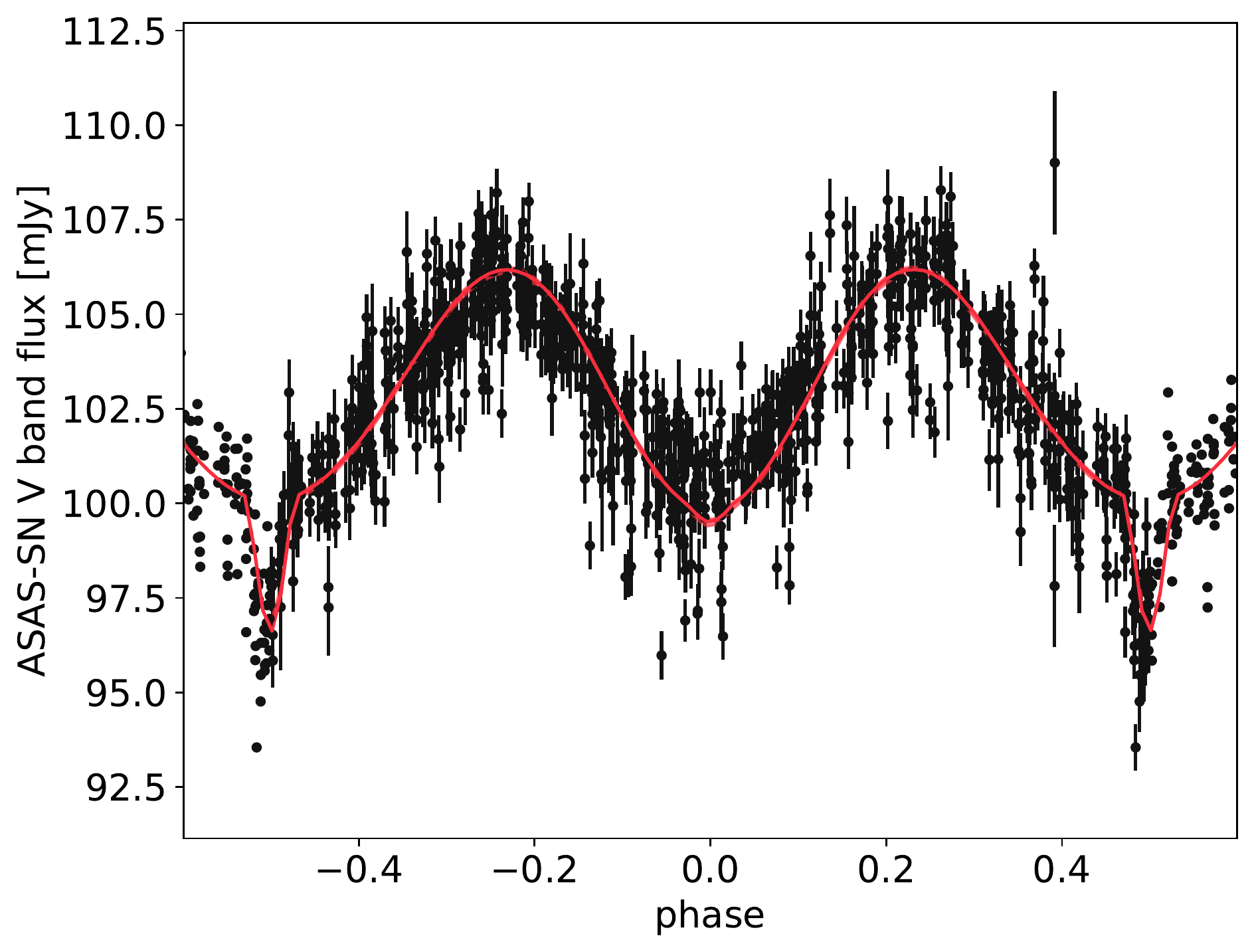}{0.5\textwidth}{}
 \fig{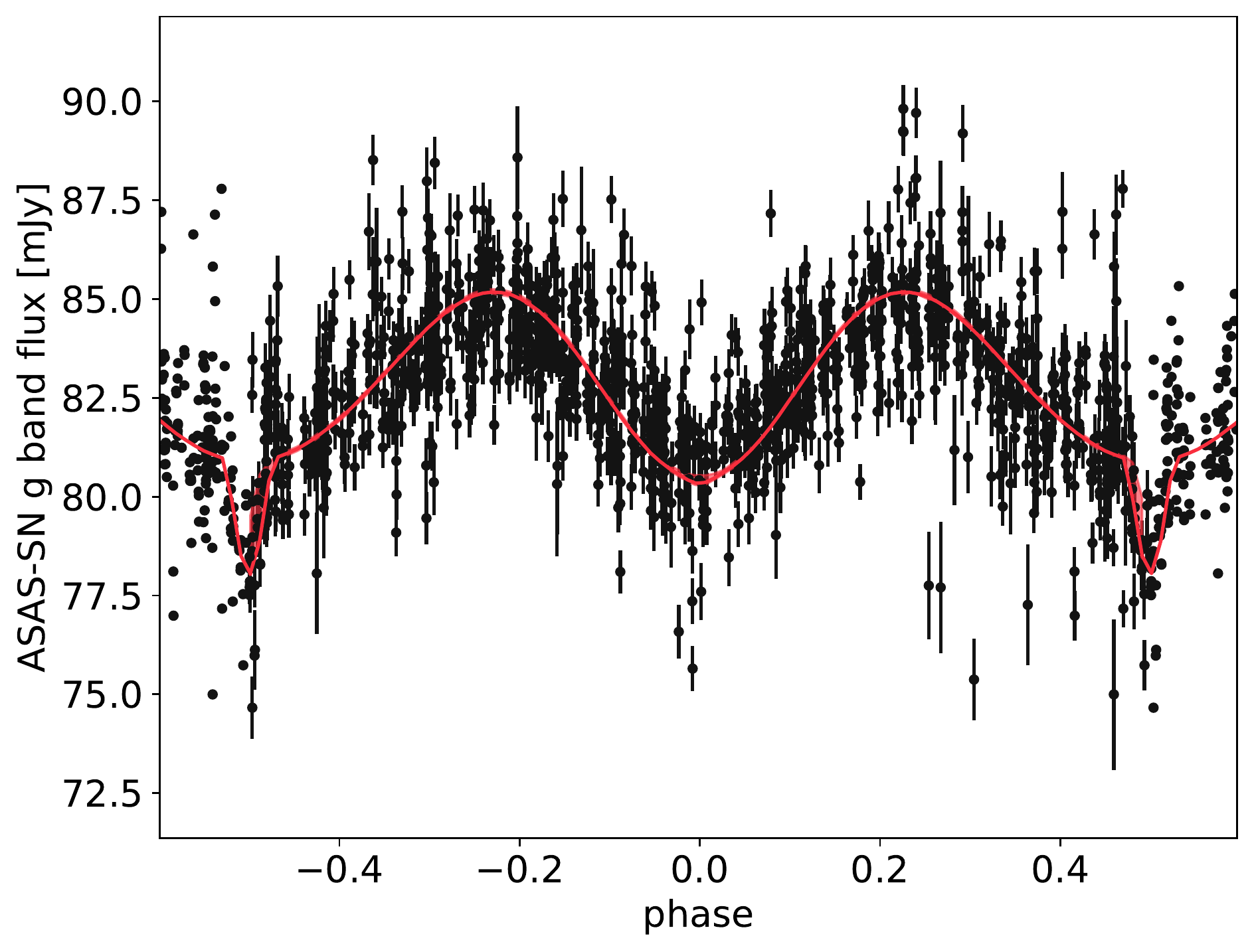}{0.5\textwidth}{}
 }
 \caption{Orbital fit of 2M1709. The top left panel shows the radial velocities derived from the APOGEE spectra of both the primary and the secondary. Remaining panels show the phase-folded light curves of the system from TESS and from ASAS-SN. The best-fit model of the orbit is overplotted on top of the data in red and blue for the primary and the secondary in the RV panel, and the model light curve is shown in red. The width of the lines and the shaded region around them show effect on the observable space from the distribution in the parameters from the emcee fit, see Sec. \ref{sec:jointfit}. The sinusoidal pattern in the light curves is attributable to the ellipsoidal variations. The strongly pronounced secondary eclipse occurs at the phase of -0.5/0.5. The primary eclipse at the phase of 0 is much weaker, only barely apparent in the TESS lightcurve.}
 \label{fig:orbit}
\end{figure*}

The system was observed with the APOGEE spectrograph \citep{wilson2010,wilson2012,majewski2017} over the course of seven epochs spanning from September 14, 2015 to March 11, 2016. APOGEE has a spectral resolution of $R\sim22,500$ and covers the wavelength range of $1.51 - 1.7 \micron$.

In the 14th data release of the Sloan Digital Sky Survey (SDSS DR14), 2M1709 was reported to have parameters of $T_{\rm eff} = 4749 \pm 80$ K and $\log g = 2.36 \pm 0.05$ dex \citep{abolfathi2018}. Using a different approach for processing on the same data reduction gave the parameters of $T_{\rm eff} = 4698.6$ K and $\log g = 3.2$ dex (without measurement uncertainty) in The Cannon \citep{ness2015}. Similarly, APOGEE Net \citep{olney2020}, also using DR14 data, report $T_{\rm eff} = 4811 \pm 81$ K and $\log g=3.426 \pm 0.009$ dex. In general, while various pipelines agree that the system is a red giant, there is a significant scatter in the derived properties, particularly \logg. It is unclear to what degree these parameters may be affected by the presence of multiple components, as they are calibrated only to single stars. However, it is clear that the signal in the near-IR spectra is dominated by a flux from a significantly cooler component than that which dominates the optical data, either from Gaia or LAMOST.

Previously, \cite{price-whelan2018} identified 2M1709 as a spectroscopic binary based on variations in APOGEE's pipeline RVs for the system. The phase coverage of the APOGEE epochs are distributed homogeneously throughout the orbit; \citet{price-whelan2018} find a unimodal orbital solution with a maximum likelihood period of 5.86562 days. This is consistent with the period of 5.86688 days that can be extracted from the eclipses in the light curve \citep{pawlak2019}.

In examining the cross-correlation function (CCF) of individual APOGEE epochs, we identified 2M1709 as a double-lined spectroscopic binary (SB2), with two components weakly resolved in five out of seven epochs. The velocities of individual components were extracted through fitting multiple Gaussians to the CCFs using GaussPy \citep{lindner2015}. These RV measurements are reported in Table \ref{tab:rv}. The cooler red giant dominates the luminosity of the APOGEE spectra resulting in a significantly stronger peak of its CCF compared to the subgiant. As the red giant is more evolved from the main sequence, and thus would have originally been the more massive of the two, we consider it to be the primary, and the subgiant to be the secondary. Despite this, the primary has a greater amplitude of oscillation in its RVs, suggesting the current mass ratio $q>5$ (Figure \ref{fig:orbit}).

\begin{deluxetable}{ccc}
\tablecaption{Radial velocity of individual components of 2M1709 \label{tab:rv}}
\tablewidth{0pt}
\tablehead{
\colhead{Date} & \colhead{$v_1$} & \colhead{$v_2$} \\
\colhead{(HJD)} & \colhead{(\kms)} & \colhead{(\kms)}
}
\startdata
57279.5853 & $98.0\pm2.4$ & $-38.0\pm7.4$ \\
57280.6039 & $68.4\pm2.6$ & $-35.3\pm10.3$ \\
57283.6088 & $-80.4\pm2.6$ & \\
57284.6336 & $32.2\pm2.9$ & \\
57454.9976 & $62.1\pm2.6$ & $-29.5\pm7.2$ \\
57458.0010 & $-95.3\pm2.7$ & $-3.5\pm8.9$ \\
57458.9648 & $-123.4\pm3.0$ & $1.0\pm12.3$ \\
\enddata
\end{deluxetable}

\subsection{ASAS-SN}
This system was identified by \cite{pawlak2019} as an eclipsing binary with $P_{\rm orb} = 5.86688$ days and a mean V magnitude of 11.44 using light curves from All-Sky Automated Survey for Supernovae \citep[ASAS-SN;][]{kochanek2017}. 2M1709 had been observed over 2597 epochs, dating from from March 2012 to February 2021. From 2012 to 2018, all epochs were recorded in V band with a zero point calibrated with APASS. In 2018, ASAS-SN brought their g-band cameras online and were able to monitor the system with both g and V bands. In examining both light curves, there do not appear to be any significant long-term trends.

\subsection{TESS}\label{sec:tess}

To date, 2M1709 has been observed by TESS in sectors 25 \& 26. We downloaded the light curve of the system using \texttt{eleanor} \citep{feinstein2019}. This package offers multiple options for detrending; \texttt{psf\_flux} was optimal for the system, as it offered the most uniform continuum flux.

However, even \texttt{psf\_flux} does not detrend the light curve perfectly. Each sector of TESS covers two orbits of the telescope, and each of the four orbits had its own distinct trend in the data. The task of removing them is made more complex due to the system's significant out-of-eclipse ellipsoidal variations. The best approach was determined to be to only take the brightest 5\% of the flux (i.e., the portion of the light curve that most closely traces the continuum) in each half-sector, fit a second-order polynomial, and normalize the light curve in that half-sector by this fit. This results in a light curve that does not have significant discrepancies across multiple orbital periods when folded (Figure \ref{fig:orbit}).

Visual examination of the TESS light curve reveals clear signatures of the system's architecture and orbital alignment. Small eclipse signatures of unequal depth are present, pointing to the difference in radii of the two components. These eclipses phase cleanly with the larger amplitude sinusoidal variability, pointing to ellipsoidal variations from an orbitally synchronized component rather than pulsation as the cause of the dominant periodic signal.

\section{System modeling}\label{sec:analysis}

\subsection{Spectral Energy Distribution (SED)}
\begin{figure}[!ht]
 \centering
 \includegraphics[width=\linewidth,trim=90 75 90 90,clip]{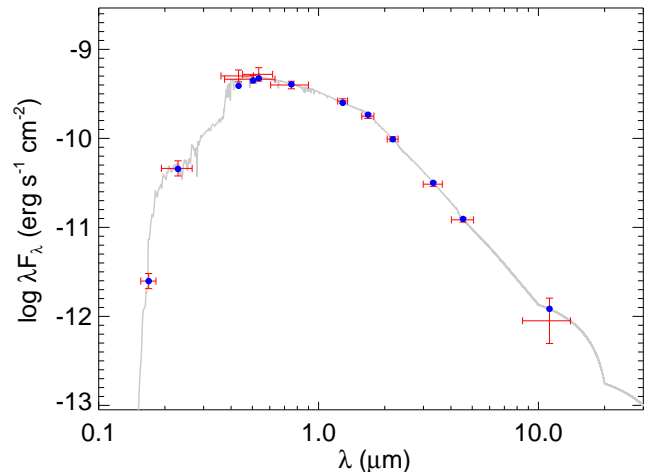}
 \caption{SED fitting of the photometry of 2M1709, using a two component fit. Vertical error bars represent the uncertainty in the flux measurements; horizontal bars represent the wavelength coverage of each filter. }
 \label{fig:sed}
\end{figure}

In order to constrain the approximate temperatures and radii of the eclipsing stars in the system, we used the methodology of \citet{stassun2016a} to fit the joint SED of the system. We used the {\it GALEX\/} FUV and NUV magnitudes, the {\it Tycho-2\/} $B_T$, $V_T$ magnitudes, the $G_{\rm RP}$ magnitudes, the {\it 2MASS\/} $J$, $H$, $K_S$ magnitudes, and the {\it WISE} $W1$--$W3$ magnitudes. Together, the available photometry spans the SED over the wavelength range 0.2 -- 10~$\mu$m (Figure~\ref{fig:sed}). 

To begin our derivation of the two effective temperatures, we considered the available spectroscopic constraints. In particular, the reported temperature estimates from optical spectra, such as LAMOST \citep[SpT = F0, $T_{\rm eff} \sim 7300~$K; $\log g = 4.15$; Fe/H = +0.07]{luo2019} and from near-IR spectra such as APOGEE \citep[$\sim$4500~K;][]{olney2020} imply one hot component and one cool component. In addition, the distance derived from the Gaia parallax of $1.014 \pm 0.015$ mas strongly constrains the total system luminosity and thus the stellar radii. Finally, we allowed the extinction ($A_V$) to be a free parameter, but limited the maximum line-of-sight extinction to the value reported at the system's position in the dust maps of \citet{Schlegel1998}. 

The resulting joint SED fit of both stars (Figure~\ref{fig:sed}) provides a good match to all of the fluxes, with a reduced $\chi^2 = 0.8$ for the following best-fit extinction, temperatures, and radii: 
$A_V = 0.13 \pm 0.06$, 
$T_1 = 4500 \pm 100$~K, 
$R_1 = 4.67 \pm 0.23$~R$_\odot$,
$T_2 = 6975 \pm 100$~K,
$R_2 = 2.95 \pm 0.15$~R$_\odot$. 
Note that the uncertainties here are formal statistical errors and are intended only to inform the priors adopted in the full solution (Sec.~\ref{sec:jointfit}). However, the fit implies that the system is composed of a red giant \citep[possibly red clump star; see][]{ting2018} and a hotter subgiant companion.

\subsection{Joint fit \label{sec:jointfit}}

\begin{figure}[!ht]
 \centering
 \includegraphics[width=\linewidth]{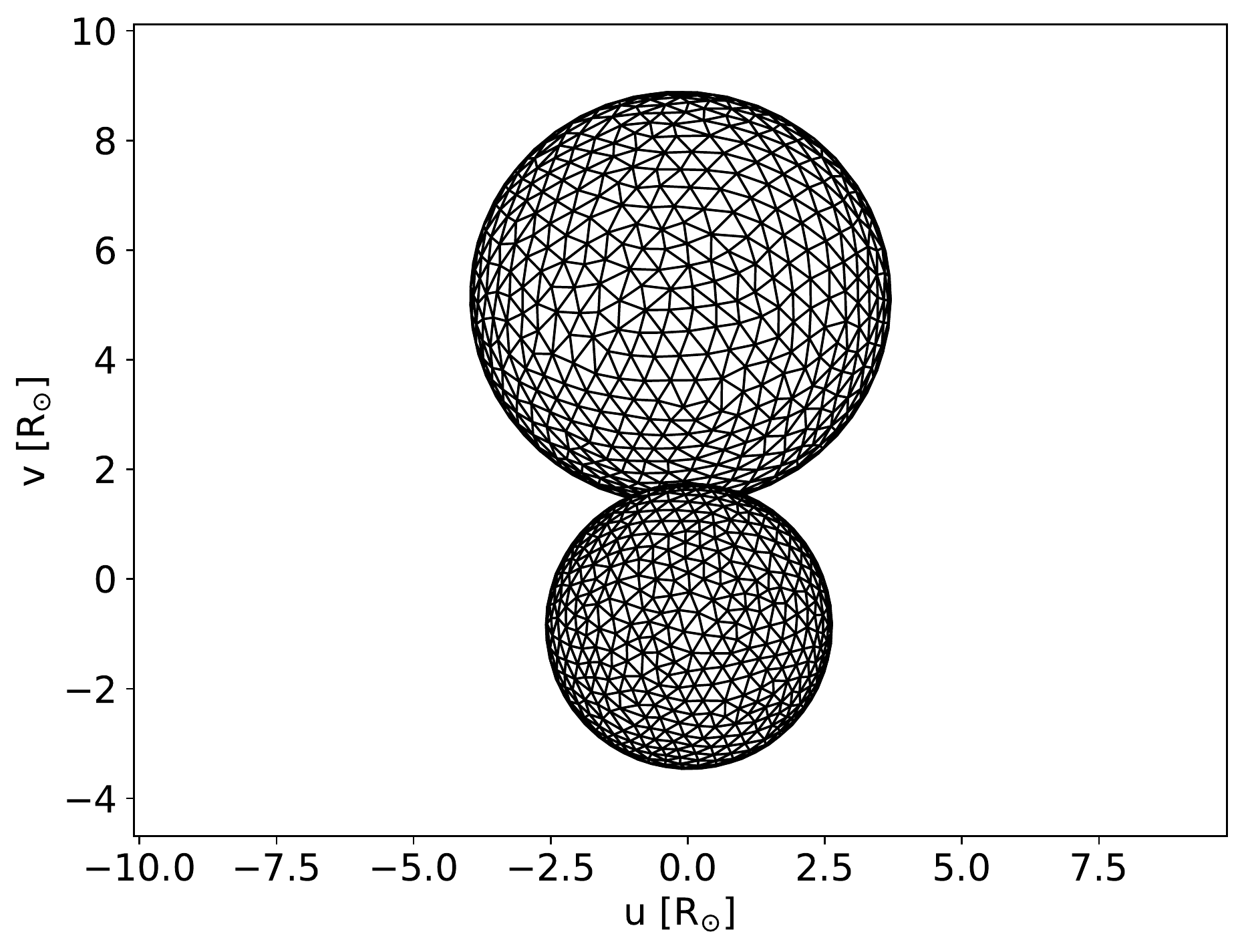}
 \includegraphics[width=\linewidth]{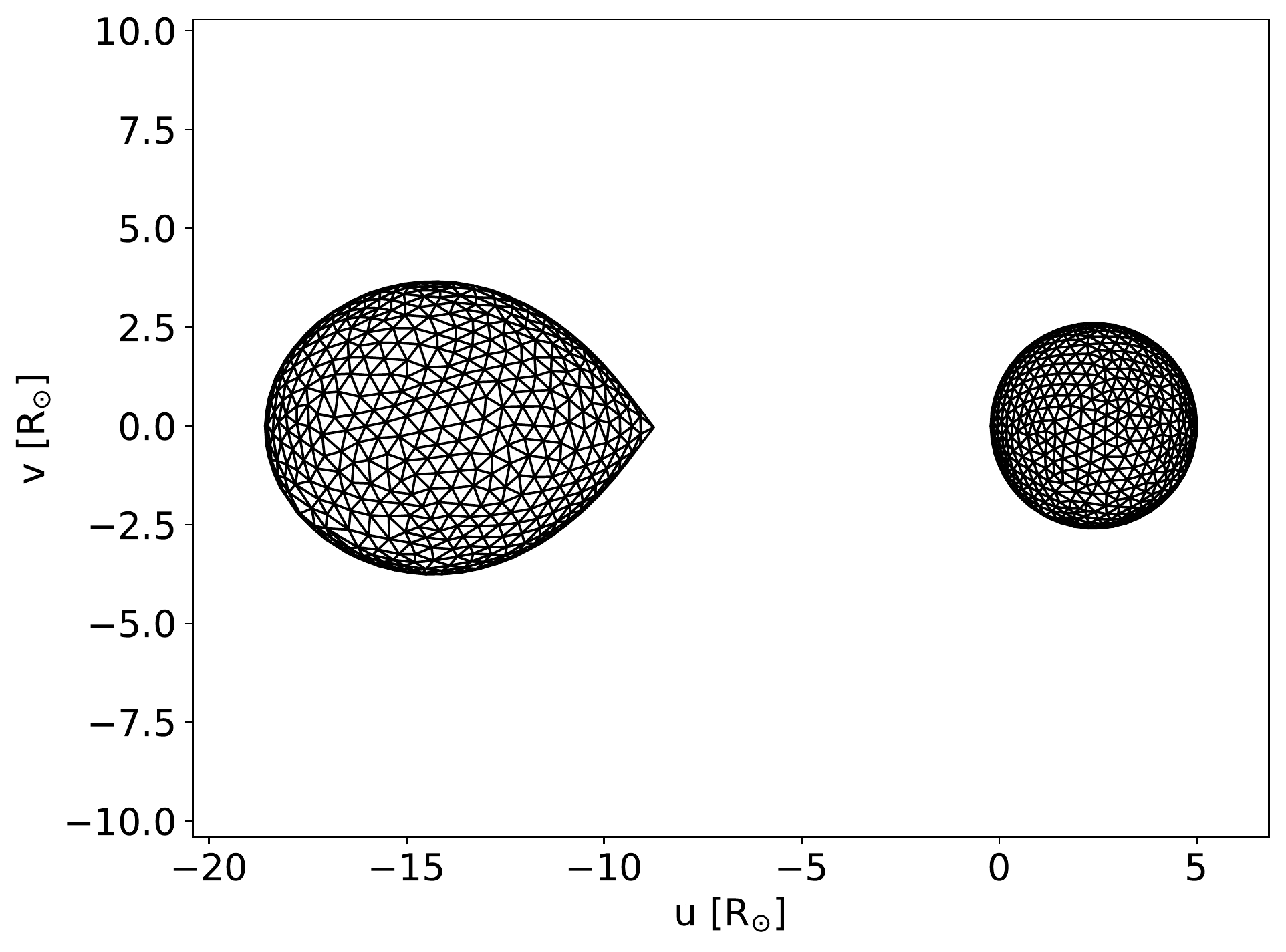}
 \includegraphics[width=\linewidth]{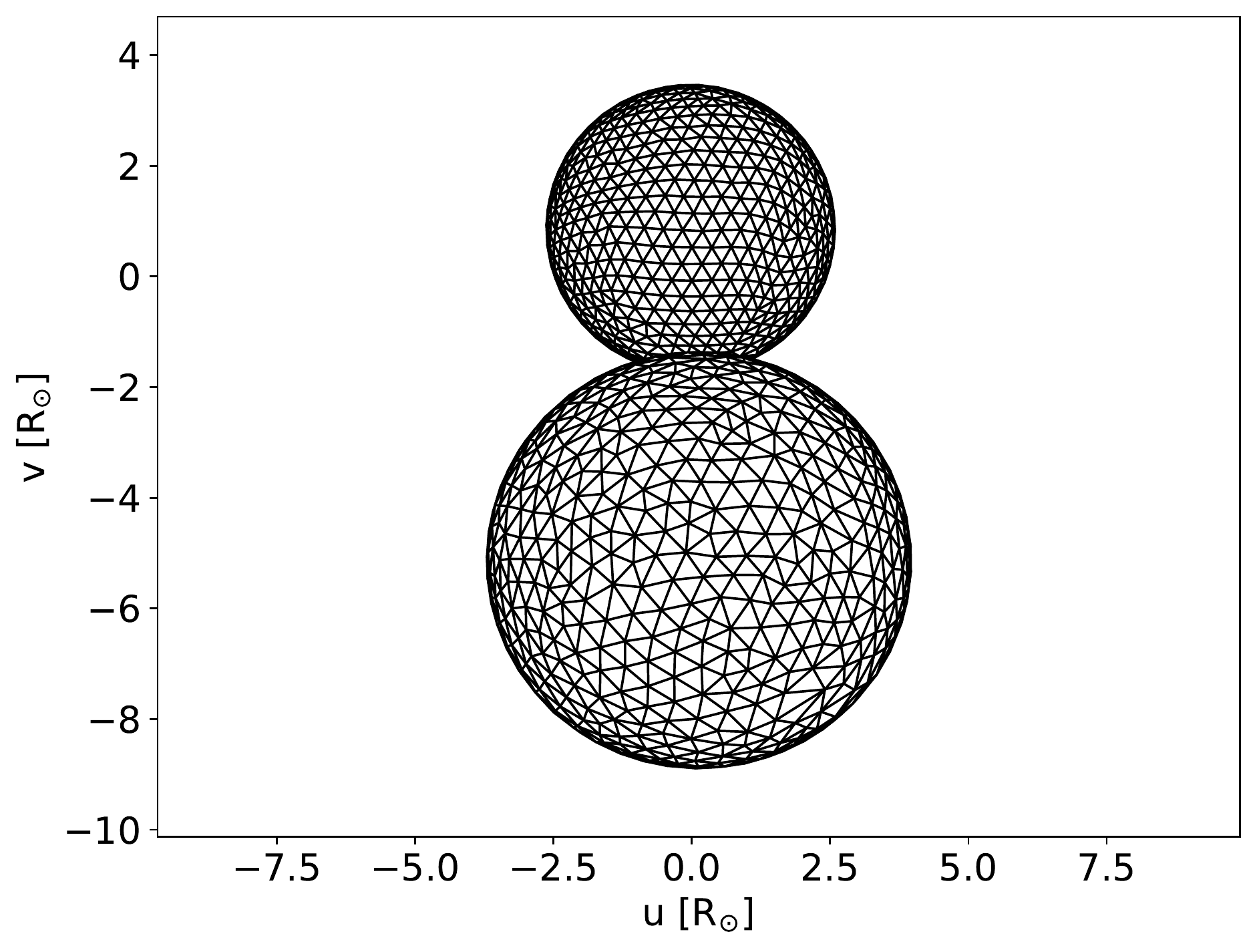}
 \caption{Mesh plots of system geometry at superior conjunction, quadrature, and inferior conjunction. }
 \label{fig:mesh}
\end{figure}

\begin{figure*}
 \centering
 \epsscale{1.2}
 \plotone{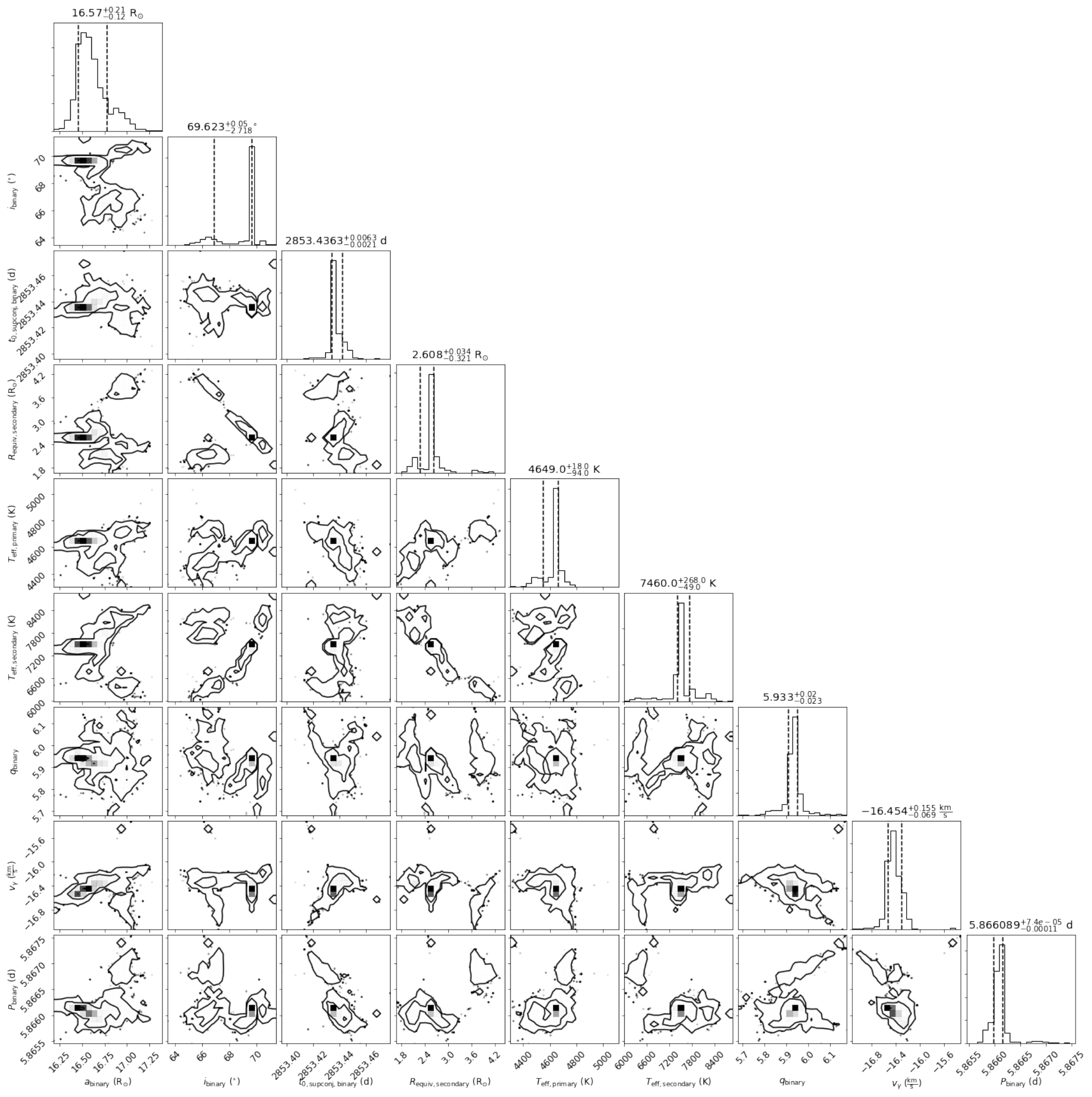}
 \caption{Corner distribution plot between the model fitting parameters}
 \label{fig:chi2_distr}
\end{figure*}

To fit the system we use PHOEBE version 2.3, which has introduced the capability of solving for the orbit from all of these data \citep{conroy2020}. In this fitting process, we have included both the light curves (TESS, as well as ASAS-SN V \& g bands) and the RVs reported in Table \ref{tab:rv}. Each star had its atmosphere set as a blackbody, with the limb darkening coefficients for each star in each band taken from \citet{claret2011}.

In addition to the inputs derived from the SED fitting, we also adopted initial parameters for the system based on an independent RV fit using The Joker \citep{price-whelan2017}. The Joker fit provided estimates of each component's mass, and the system's mass ratio and semi-major axis, along with uncertainties that could be used as the initial distributions for the subsequent PHOEBE orbital fitting. We then iterated on these inputs with PHOEBE, both manually, and through using various solvers, to produce an acceptable $\chi$-by-eye fit of all of the data; the resulting parameter estimates are listed in the 'Initial Estimate' column of Table \ref{tab:orbit}. In exploring the parameter space, the primary star typically overfilled its Roche lobe. Thus, the system was set to be semi-detached, which fixed $R_1$ to the size of the Roche lobe.

Finally, we used the emcee sampler built into PHOEBE to refine the fit and to estimate the uncertainties. We initialized chains using Gaussian distributions on 12 parameters: period ($P_{\rm orb}$), inclination ($i$), semimajor axis ($a$), radius of the secondary star ($R_2$), mass ratio ($q$), central velocity of the system ($\gamma$), time of passage of the superior conjunction ($T_0$), and effective temperatures of both stars ($T_{\rm eff{1,2}}$). The widths of the Gaussian distributions were approximated from the earlier exploration of the parameter space with PHOEBE, as well as the independent RV fit from the Joker. Additionally, we fit three ``nuisance'' parameters corresponding to the bandpass luminosity of each band for which light curve is available - they are recommended by PHOEBE to avoid underestimating the scatter when running the emcee solver, although they are not necessarily physically meaningful, especially for the normalized light curves (such as TESS).

We initialized emcee with 250 walkers, each of which was computed for 2000 steps. The burn-in was set to 600, which was the point where most of the walkers have converged in examining the $\ln(\mathrm{probability})$ plot, and walkers with $\ln(\mathrm{probability})<-5e5$ were rejected. The uncertainties were set using the 68\% percentile distribution of the scatter of the resulting chains, and they were propagated to the fixed parameters that depended on them, such as $R_1$ and $M_{1,2}$. The fitted parameters are listed in Table \ref{tab:orbit}. The 3-dimensional model showing the best-fit geometry of the two stars is shown in Figure \ref{fig:mesh}.
The corner plot showing the covariances between model parameters is shown in Figure \ref{fig:chi2_distr}.

There are slight differences between the derived properties of the star in the MCMC from PHOEBE and in the initial RV only fit by the Joker. The RV amplitudes derived by the Joker for the two stars are $K1=114.8 \pm 1.3$ \kms, and $K2=19.5 \pm 4.5$ \kms. Assuming $i\sim69^\circ$, as derived independently from the light curve analysis, these velocities imply $M_1=0.27\pm0.03$ \msun, $M_2=1.57\pm0.17$ \msun, and $a=16.8\pm0.6$ \rsun. On the other hand, the fit from PHOEBE estimates $M_1=0.256^{+0.0098}_{-0.0059}$ \msun, $M_2=1.518^{+0.057}_{-0.031}$ \msun, and $a=16.57^{+0.21}_{-0.12}$ \rsun. That is to say, both fits are comparable with each other, within the errors, but the uncertainties from PHOEBE are typically $\sim$ 3 times smaller than in the RV-only fit. As PHOEBE does also simultaneously fit the light curves, it is possible that the additional data may help in better constraining the solution of these parameters (not the least of which is the period, as the light curves span a longer baseline). It is also possible that the uncertainties from PHOEBE may be underestimated by a factor of 3.

\begin{deluxetable*}{cccccc}
\tablecaption{Parameters of 2M1709\label{tab:orbit}}
\tablewidth{0pt}
\tablehead{
\colhead{Parameter} & \colhead{RV-only fit} &\colhead{Initial estimate} &\colhead{Distribution width} & \colhead{Solved Value} & \colhead{Model Value}\\
\colhead{} & \colhead{(The Joker)} & \colhead{(PHOEBE)} & \colhead{(PHOEBE)} & \colhead{(PHOEBE)} & \colhead{(MESA)} }
\startdata
Period (day) & $5.8661 \pm 0.0005$ &5.866238& 0.001\tablenotemark{$^b$}& $5.866089^{+7.4e-05}_{-0.00011}$ & 5.866089\\
$q$ & $5.88 \pm 0.33$ & 5.88 & 0.33\tablenotemark{$^c$} & $5.933^{+0.02}_{-0.023}$ & 5.923 \\
$\gamma$ (\kms) & $-16.1 \pm 0.5$ & $-16.136$ & 0.5\tablenotemark{$^c$}& $-16.454^{+0.155}_{-0.069}$ & \\
$a$ (\rsun) & $16.8 \pm 0.6$\tablenotemark{$^a$}&16.8& 0.6\tablenotemark{$^c$} & $16.57^{+0.21}_{-0.12}$ & 16.75\\
$M_1$ (\msun) &$0.27 \pm 0.03$\tablenotemark{$^a$}&&& $0.256^{+0.0098}_{-0.0059}$ & 0.256 \\
$M_2$ (\msun) &$1.57 \pm 0.17$\tablenotemark{$^a$}&&& $1.518^{+0.057}_{-0.031}$ & 1.568 \\
$R_1$ (\rsun) &&&& $3.961^{+0.049}_{-0.032}$ & 3.963\\
$R_2$ (\rsun) &&2.5 &1.0\tablenotemark{$^b$} & 2.608$^{+0.034}_{-0.321}$ & 2.325\\
$i$ (deg) && 69& 5\tablenotemark{$^b$} & 69.623$^{+0.05}_{-2.718}$ & \\
$T_{\rm eff,1}$ (K) &&4500&1000\tablenotemark{$^b$}& $4649^{+18}_{-94} $ & 4674 \\
$T_{\rm eff,2}$ (K)& &7300 &1000\tablenotemark{$^b$}& $7460^{+268}_{-49}$ & 7412\\
$T_0$ (JD) && 2458853.4202 &0.1\tablenotemark{$^b$}& 2458853.4363$^{+0.0063}_{-0.0021}$ & 
\enddata
\tablenotetext{a}{Assuming $i\sim69^\circ$ from an independent PHOEBE fit to explore the system's geometry.}
\tablenotetext{b}{Distribution width set to arbitrarily large values that are outside of plausible bounds in the initial exploration of the dataset}
\tablenotetext{c}{Set from the uncertainties in the RV-only fit by the Joker.}
\end{deluxetable*}

The resulting fits to the input light curves and RV measurements are shown in Figure \ref{fig:orbit}. The meshes showing the shapes and orbital separation of the two stars are shown in Figure \ref{fig:mesh}. The configuration of the system is such that the distortion in the shape of the primary (albeit less massive) star are strongly apparent. The eclipses are grazing, with the two stars only barely appearing to pass in front of the other, resulting in a very weak eclipse signature (especially when the hotter secondary is passing in front of the cooler primary).

\section{Evolutionary modeling}
 \label{sec:mesa}

To understand the progenitor binary and predict the future of this system, we have constructed detailed stellar evolution models that include MT with the Modules for Experiments in Stellar Astrophysics (MESA; Version 12115; \citealt{2011ApJS..192....3P,2013ApJS..208....4P,2015ApJS..220...15P,2018ApJS..234...34P,2019ApJS..243...10P}) package.\footnote{ Our inlist is shared at http://doi.org/10.5281/zenodo.5108975.} The two stars are evolved simultaneously in the simulation. We adopt $Z = 0.01$ as the initial metallicity (as both APOGEE and LAMOST favor sub-solar values $\sim$ -0.3 dex). Magnetic braking (with $\gamma=4$, where $\gamma$ sets the strength of the magnetic braking), and mass loss from the system are included and affect the system angular momentum during the binary evolution. Gravitational wave radiation is also included, but does not significantly impact the system angular momentum. We do not consider tidal effects because at this short period and gigayear age the system will be synchronized.

Our goal was to produce a system matching the observations in the mass - orbital period plane as well as the radius - $T_{\rm eff}$ plane. We call the resulting model the best-fit (BF) model.

We note that although there is a general agreement between the BF model and the observations, there may be a slight difference in the definition of radius between MESA and PHOEBE. MESA uses the photometric radius at which $T$ of a star is equal to \teff, given a particular atmospheric model, assuming a spherical geometry. On the other hand, PHOEBE can account for a non-spherical geometry, and it defines the equivalent radius as a radius of a sphere that would contain the volume of the star. If a star is a sphere (as is the case with the subgiant), the two radii are likely to be comparable. However, as the red giant donor star has filled up its Roche lobe, the differences may be more pronounced -- but, at most, they are expected to be $<$10\%.

The initially searched parameter space included the initial orbital period ranging from 1 to 10 days with a step of 1 day, initial donor from 1.0 $M_{\odot}$ to 1.5 $M_{\odot}$ with a step of 0.1 $M_{\odot}$, the accretor masses slightly smaller than the donor mass, and MT efficiency from 10\% to 70\%. Once we found the grid point that was matching closest to the data, the grid is narrowed down around it with smaller spacing to find the BF model. The system's measured present-day masses, radii, orbital periods and $T_{\rm eff}$ values are reproduced with initial donor and accretor masses of $M_{\rm d}=1.2\,M_{\odot}$ and $M_{\rm a}=1.11\,M_{\odot}$, respectively, and an initial $P_{\rm orb}=3.43$ days. The required MT efficiency derived by the best fit is 49\%, which means 49\% of the change in donor mass is transferred to the accretor star. In 1-D stellar evolution simulation, the MT efficiency is set as a constant during the MT.

This solution is a unique minimum. Several constraints drive the solution towards it. A donor star with a larger initial mass ($M_{\rm d}>1.2\,M_{\odot}$) would produce a red giant that is hotter than the observations indicate. On the other hand, a less massive donor star cannot ever reach the observed radius of the red giant at $P_{\rm orb}=5.866$ days. The accretor star produces the subgiant, also an evolved star in the system. The chosen accretor mass is slightly less than that of the donor star, ensuring that the accretor does not evolve off the main sequence before the donor, and also ensuring that the accretor will evolve off the main sequence while the donor is ascending the red giant branch. The initial $P_{\rm orb}$ is constrained by the observed $P_{\rm orb}$, and the the angular momentum within the system, (which is computed given the stellar masses, and later evolved given the physics implemented within MESA). A longer initial $P_{\rm orb}$ results in a wider system at the present day. A smaller initial $P_{\rm orb}$ results in a contact binary before the two stars reach their observed masses. Finally the observed accretor mass determines the MT efficiency. In the simulation, the radius is not an initial parameter, the code finds the stellar structure that satisfies the hydrostatic equilibrium at each timestep. The radius is an output. As we chose a proper metallicity and the mixing length factor, once the masses agree with the observation, the radii are matched as well.

\begin{figure}[tp]
	\includegraphics[width=0.5\textwidth]{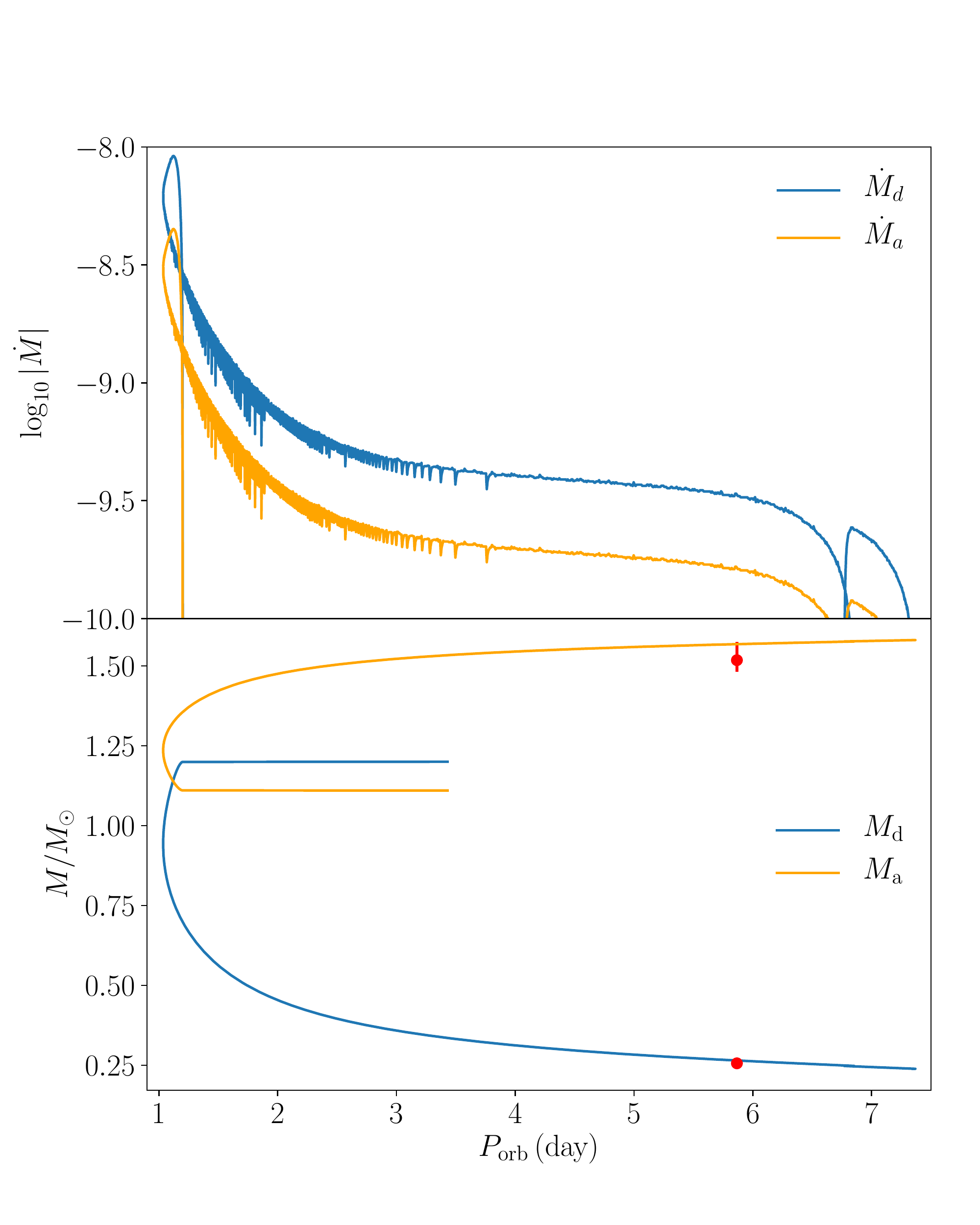}
	\caption{The upper plot shows $\dot{M}$ as a function of $P_{\rm orb}$ for the donor star (blue) and the accretor star (orange), with an initial $P_{\rm orb}=3.43$ days, initial $M_{\rm d}=1.2\,M_{\odot}$ and initial $M_{\rm a}=1.11\,M_{\odot}$. The MT efficiency is fixed at 49\%. The small scale spikes along the curve are driven by the numerical artefacts.
	The bottom panel shows the mass of the donor star (blue) and accretor star (orange) during the evolution. The red dots mark the current $P_{\rm orb}=5.866$; by construction the corresponding masses of the two stars ($0.256\,M_{\odot}$ and $1.518\,M_{\odot}$) agree closely with their measured masses. The uncertainty of the accretor mass at $P_{\rm orb}=5.866$ is also shown. The errorbar of the donor star at $P_{\rm orb}=5.866$ is smaller than the dot size so it is not shown on the bottom panel. }
	\label{fig:Mdot_P.pdf}
\end{figure}

The two panels of Figure \ref{fig:Mdot_P.pdf} show the MT rate $\dot{M}$ and the masses of the two stars as a function of $P_{\rm orb}$, which first shrinks and then grows as the system ages. The MT launches at $P_{\rm orb}=1.22$ days, when the system age is 4.58 Gyr. The MT rate, which is proportional to $\exp(R/R_{\rm RL})$, where $R$ is the radius of the donor star given its mass at a given evolutionary state, and $R_{\rm RL}$ is the radius of the Roche-lobe for the donor. Early on, MT keeps increasing due to the increase in the ratio between the donor star's radius and the donor star's Roche-lobe radius (a function of the mass ratio and the system separation \citep{1983ApJ...268..368E}); it decreases at later timesteps as the mass of the donor decreases. The MT is non-conservative; 51\% of the mass change of the donor star escapes from the system. The system separation continues to shrink due to magnetic braking and this mass loss until $P_{\rm orb}$ reaches 1.04 days. At the minimum $P_{\rm orb}$, where the system age is 4.62 Gyr, the donor and accretor masses are $0.94\,M_{\odot}$ and $1.24\,M_{\odot}$, respectively. The model reaches the current $P_{\rm orb}$ and the measured masses of the two stars at 5.26 Gyr. At 5.49 Gyr and $P_{\rm orb}=7.35$ days, the MT stops and the system becomes a pre-WD and a blue straggler pair. 

To make an evolutionary track pass exactly through the current accretor mass (the red dot in the bottom panel of Figure \ref{fig:Mdot_P.pdf}) at the current period, the MT efficiency needs to be 44\%, slightly smaller than in the BF model. All the other initial parameters are the same as the BF model and they yield the same evolutionary story.

\begin{figure}[tp]
	\includegraphics[width=0.5\textwidth]{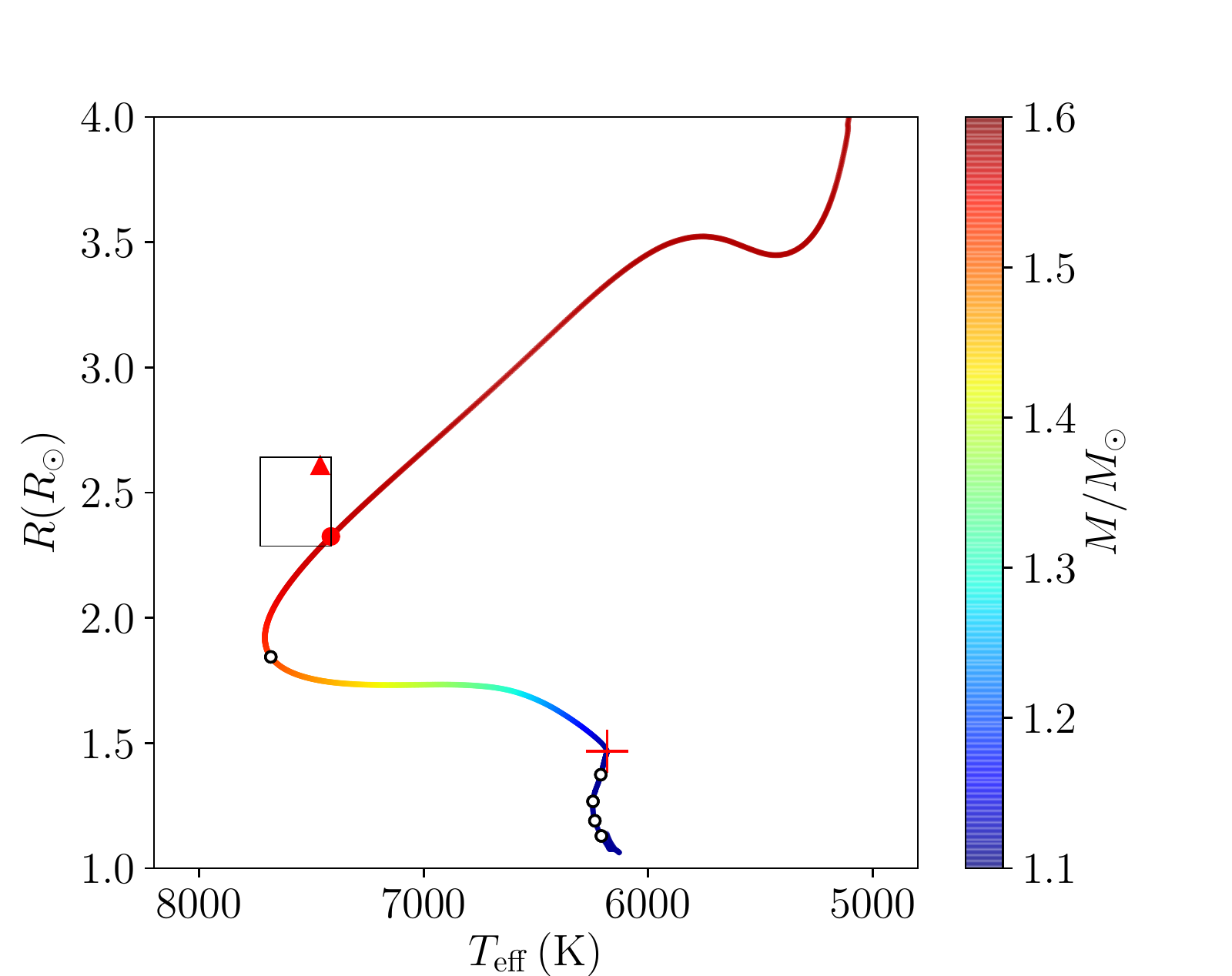}
	\caption{Evolutionary track of the accretor star in the radius-effective temperature plane from the BF model. The color gives the accretor's mass. The open circles are spaced at every 1 Gyr. The red cross shows the onset of mass transfer. The red triangles represent the observed values from PHOEBE fitting. The size of the box around the red star indicates the 68\% percentile uncertainties. The red dot on the evolutionary track marks the current $P_{\rm orb}=5.866$, and correspond to the same red dot on the orange line of Figure \ref{fig:Mdot_P.pdf}.}
	\label{fig:R_Teff2_comp.pdf}
\end{figure}

\begin{figure}[tp]
	\includegraphics[width=0.5\textwidth]{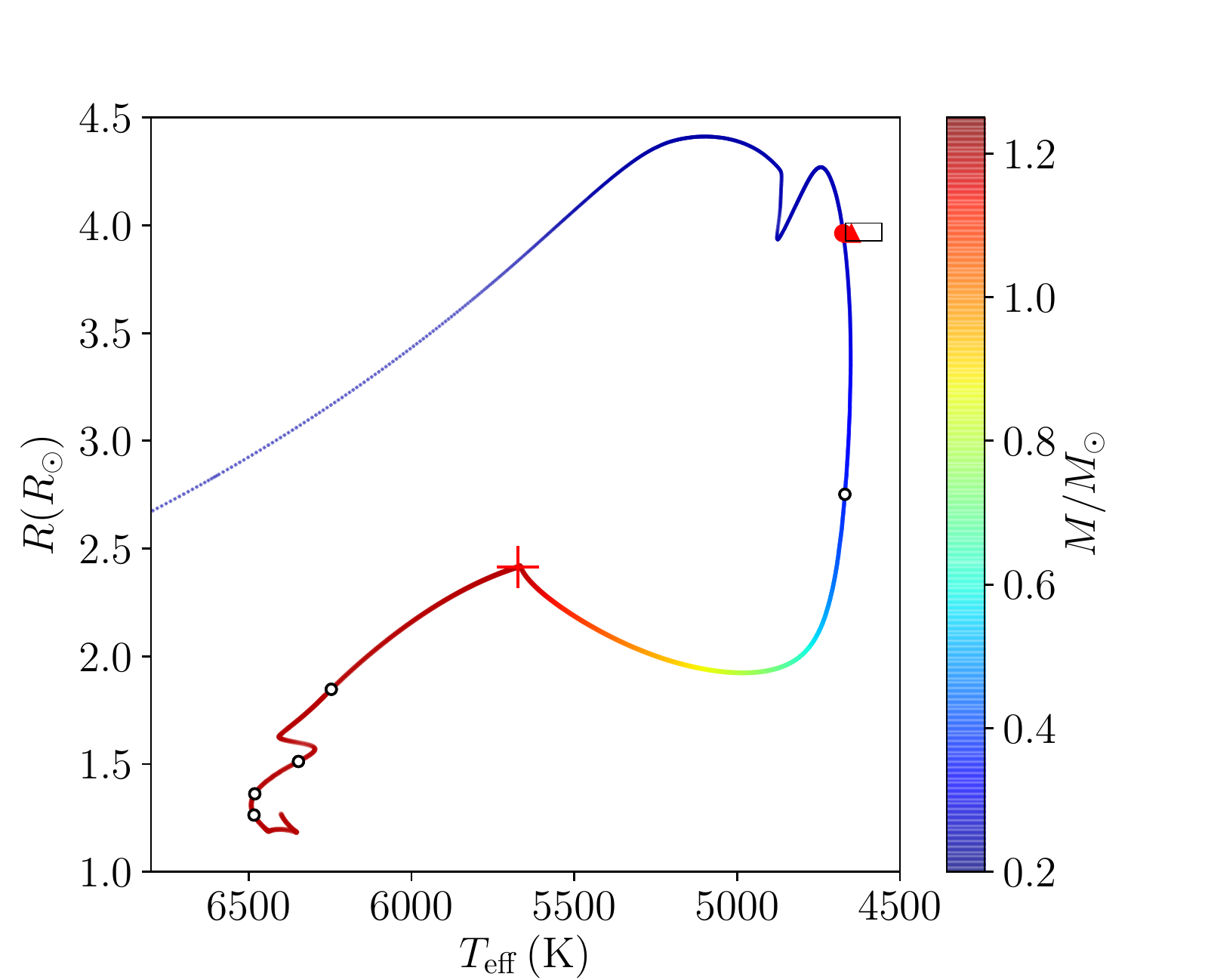}
	\caption{Evolutionary track of the donor star in the radius-effective temperature plane from the BF model. The red triangle represents the observed values from PHOEBE fitting. Other figure properties as in Figure \ref{fig:R_Teff2_comp.pdf}. The red dot corresponds to the same model as the red dot on the blue line of Figure \ref{fig:Mdot_P.pdf}.}
	\label{fig:R_Teff.pdf}
\end{figure}

Figure \ref{fig:R_Teff2_comp.pdf} displays the evolutionary track of the accretor star in the BF model from the main sequence through the early red giant phase. The zero-age-main-sequence position of the accretor star is near $(T_{\rm eff},R)=(6170\,{\rm K}, 1.08\,R_{\odot})$. 
The mass accretion starts at $(T_{\rm eff},R)=(6183\,{\rm K},1.47\,R_{\odot})$; this moment is marked with a red cross in both Figures \ref{fig:R_Teff2_comp.pdf} and \ref{fig:R_Teff.pdf}. The mass accretion rate is always slow ($M_{\odot}< 10^{-8} M_{\odot}$/yr, see Figure \ref{fig:Mdot_P.pdf}), so the accretor has time to adjust to new hydrostatic and thermal equilibria. Thus the effective temperature and luminosity of the accretor keeps increasing with the increasing mass of the star. When the nuclear activity switches from core burning to shell burning, the effective temperature begins to drop and the accretor enters the subgiant phase. During the subgiant phase, the accretor increases in radius and luminosity and decreases in effective temperature. 
The red dot on Figure \ref{fig:R_Teff2_comp.pdf} represents the same model as the red dot on the orange line of Figure \ref{fig:Mdot_P.pdf}, which successfully reproduces the observed $P_{\rm orb}=5.866$ days and both stellar masses. 

In the radius - $T_{\rm eff}$ plane, the BF model underestimates the maximum likelihood accretor radius of the PHOEBE light curve analysis. While the accretor radius in the BF MESA model is consistent with the notably asymmetric uncertainties in the PHOEBE-derived radius, we explored changes in the MESA model that could potentially relieve this tension. The radius difference mainly derives from the model accretor mass. A smaller mass results in a smaller radius in the subgiant phase. Thus a more massive accretor star could reproduce the accretor's larger observed present-day radius.

To examine scenarios that would generate a more massive present-day accretor, we explored an alternative model with higher mass transfer-efficiency, which was able to match the PHOEBE results in the radius - $T_{\rm eff}$ plane. In this model, we took the initial $M_{\rm d}$ value to be the same as the BF model. We find that an initial $M_{\rm a}=1.10M_{\odot}$, $P_{\rm orb}=3.44$ and mass-transfer-efficiency of 60\% reproduce the radius and effective temperature of our PHOEBE analysis at the observed orbital period.
However, this model cannot reproduce the accretor's present location in the mass - $P_{\rm orb}$ plane. At $P_{\rm orb}=5.866$ days (the system's present-day period), the accretor mass asymptotes towards $1.66\,M_{\odot}$ (i.e., $\sim$10\% higher than the accretor's measured present-day mass). 

Focusing on the accretor's end-state in this alternate model, at $P_{\rm orb}=7.39$ days, the accretor becomes a red giant star and fills its Roche-lobe radius. The system then forms a contact binary and the simulation is terminated by MESA.

Figure \ref{fig:R_Teff.pdf} gives the evolutionary track for the donor star in the BF model. Similar to Figure \ref{fig:R_Teff2_comp.pdf}, open circles are placed at every 1 Gyr and the red cross gives the onset of the mass transfer phase. The zero-age-main-sequence position of the donor star is near $(T_{\rm eff},R)=(6437\,{\rm K},1.19\,R_{\odot})$. At 3.41 Gyr, the donor star develops a helium core, which corresponds to the blue hook after the third open circle. As the magnetic braking keeps shrinking the orbit and the donor star increases in radius, mass transfer launches at $(T_{\rm eff},R)=(5673\,{\rm K},2.42\,R_{\odot})$, where the helium core mass of the donor star is 0.12 $M_{\odot}$. Therefore, the evolution of this system is Case B mass transfer. Examination of the system timescales show that the mass transfer is thermally driven. At the donor star's present-day observed properties the system is still in the MT phase. The mass loss rates of the donor star are from $10^{-9}$ to $10^{-10}\,\dot{M}_{\odot}/{\rm yr}$ (Figure \ref{fig:Mdot_P.pdf}). The simulation ends with a pre-WD and blue straggler pair after the ceasing of the MT.

Changing the initial accretor mass and the MT efficiency has only a minor impact on the evolution of the donor star. The variations explored above for the accretor star produce a very similar donor evolutionary track.

One of the reasons that we believe the system is still in the MT phase is the mass-period relation for post-mass transfer systems. \citet{1999A&A...350..928T} gave this relation when running detailed binary evolution models for the millisecond pulsars. This relation has been since validated by the discovery of hundreds of other close binary systems (\citealt{1999A&A...350..928T,2014ApJ...781L..13T}). Our system currently has a donor star with a mass of 0.26 $M_{\odot}$. If MT were now completed, the mass-period relation for post-mass transfer systems indicates that the corresponding $P_{\rm orb}$ should be near 20 days, much larger than the measured current period of 5.866 days. The fact that this system does not agree with the known mass-period relation for post-mass transfer systems, along with the mass transfer rate implied by our MESA modeling, argues that the MT is ongoing. Importantly, at the completion of MT the BF model predicts an orbital period of $P_{\rm orb}=7.35$ days and a donor mass of 0.24 $M_{\odot}$, which is consistent with the mass-period relation.

To summarize the system evolution, initially the system shrinks in separation, mainly due to magnetic braking. Then, with the onset of hydrogen shell burning and the consequent growth in donor radius, MT launches. The MT is non-conservative. After the mass ratio is reversed, the system begins to expand. With completion of the MT due to the exhaustion of the shell hydrogen in the donor star, the system will yield a blue straggler-white dwarf system.

\section{Summary and Conclusions}\label{sec:summary}

2M1709 is a post-mass transfer binary system in a $P_{\rm orb}=5.87$ day grazing eclipsing orbit that is undergoing mass transfer. Through analyzing the system's SED and through performing a joint modelling of TESS and ASAS-SN light curves, and APOGEE radial-velocity measurements, we measure the system's present-day properties, which consist of an $M = 0.26 \,M_{\odot}$ evolved giant / donor star and an $M = 1.52 \,M_{\odot}$ subgiant / accretor, which, through gaining mass from the giant, has become a blue straggler.

To understand the formation history and likely evolutionary path of this system, we have performed detailed modelling of the system using the MESA binary evolution code. In particular, we searched for an initial main-sequence binary configuration that can reproduce the system's present-day masses, temperatures, radii, and orbital period. We require the two components to be coeval, but allow the mass transfer efficiency to remain a free parameter for optimizing the fit. The model that best fits the present day system begins with a binary with a 1.2 $M_{\odot}$ primary and 1.11 $M_{\odot}$ secondary in a closer $P_{\rm orb}$ = 3.4 day orbit. This model indicates that at an age of $\sim$ 5 Gyrs, the primary evolved onto the red giant branch, filled its Roche-lobe and initiated mass transfer onto the secondary component. Of the mass lost by the primary, 49\% ultimately was accreted by the secondary -- the remainder was lost, increasing the system's specific angular momentum such that the system's orbit expanded and its period increased to its current $P_{\rm orb}=5.87$ day configuration. 

The model indicates that the system remains in a mass transfer configuration currently, with the donor contributing $\dot{M} \sim $10$^{-9}$ M$_{\odot}/{\rm yr}$.

\software{PHOEBE \citep{conroy2020}, GaussPy \citep{lindner2015}, The Joker \citep{price-whelan2017}, eleanor \citep{feinstein2019}, TOPCAT \citep{topcat} MESA; Version 12115; \citep{2011ApJS..192....3P,2013ApJS..208....4P,2015ApJS..220...15P,2018ApJS..234...34P,2019ApJS..243...10P} }

\begin{acknowledgements}

We thank Nancy Evans for useful advice regarding the Cepheid classification of this source, and Kyle Conroy for assisting with the initialization of PHOEBE. We sincerely thank the anonymous referee for a prompt and helpful report that improved the presentation of our results.\\
M.K. and K.C. acknowledge support provided by the NSF through grant AST-1449476; A. M. acknowledges additional support from NASA Award NNX15AJ98H, under the Washington NASA Space Grant Consortium and Chandra Award Number GO9-20006X issued by the Chandra X-ray Center, which is operated by the Smithsonian Astrophysical Observatory for and on behalf of the National Aeronautics Space Administration under contract NAS8-03060. M.S. and R.D.M. acknowledge funding support from NSF AST-1714506 and the Wisconsin Alumni Research Foundation.

\end{acknowledgements}

\bibliographystyle{aasjournal}
\bibliography{ms.bbl}{}

\end{CJK*}
\end{document}